\documentclass[aps,prbbib,twocolumn,epsf]{revtex4}

\usepackage{graphicx}
\usepackage[usenames]{color}

\usepackage{natbib}

\begin{document}
\draft

\title{Interfacial magnetic anisotropy from a 3-dimensional Rashba substrate}
\author{Junwen Li$^{1,2}$, Paul M. Haney$^1$}
\affiliation{1.  Center for Nanoscale Science and Technology, National Institute of Standards and Technology, Gaithersburg, MD 20899 \\
2.  Maryland NanoCenter, University of Maryland, College Park, MD 20742, USA \\
}
\begin{abstract}
We study the magnetic anisotropy which arises at the interface between a thin film ferromagnet and a 3-d Rashba material.  The 3-d Rashba material is characterized by the spin-orbit strength $\alpha$ and the direction of broken bulk inversion symmetry $\hat n$.  We find an in-plane uniaxial anisotropy in the $\hat{z}\times\hat{n}$ direction, where $\hat z$ is the interface normal.  For realistic values of $\alpha$, the uniaxial anisotropy is of a similar order of magnitude as the bulk magnetocrystalline anisotropy.  Evaluating the uniaxial anisotropy for a simplified model in 1-d shows that for small band filling, the in-plane easy axis anisotropy scales as $\alpha^4$ and results from a twisted exchange interaction between the spins in the 3-d Rashba material and the ferromagnet. For a ferroelectric 3-d Rashba material, $\hat n$ can be controlled with an electric field, and we propose that the interfacial magnetic anisotropy could provide a mechanism for electrical control of the magnetic orientation.
\end{abstract}

\maketitle

\section{Introduction}

Interfacial magnetic anisotropy plays a key role in thin film ferromagnetism.  For ultra thin magnetic layers (less than 1 nm thickness), the reduced symmetry at the interface and orbital hybridization between the ferromagnet and substrate can lead to perpendicular magnetic anisotropy \cite{carcia1985perpendicular,yang2011first,daalderop1994magnetic}.  Perpendicular magnetization in magnetic multilayers can enable current-induced magnetic switching at lower current densities \cite{ikeda2010perpendicular,mangin2006current}.  Interfacial magnetic anisotropy is also at the heart of several schemes of electric-field based magnetic switching.  In this case an externally applied field can modify the electronic properties of the interface, changing the magnetic anisotropy and leading to efficient switching of the magnetic layer \cite{wang2012electric,li2015thermally,maruyama2009large,shiota2012induction}.  The combination of symmetry breaking at the interface and the materials' spin-orbit coupling generally leads to an effective Rashba-like interaction acting on the orbitals at the interface \cite{park2013orbital,haney2013}.  The interfacial magnetic anisotropy can be studied in terms of a minimal model containing both ferromagnetism and Rashba spin-orbit coupling \cite{barnes2014rashba}.  The interfacial magnetic anisotropy direction is a structural property of the sample leading to easy- or hard-axis out-of-plane anisotropy, and isotropic in-plane anisotropy.

There has been recent interest in materials with strong spin-orbit coupling which lack structure inversion symmetry in the bulk.  These are known as 3-d Rashba materials, and examples include BiTeI \cite{ishizaka2011giant} and GeTe \cite{pawley1966diatomic}.  In BiTeI, the structure inversion asymmetry results from the asymmetric stacking of Bi, Te, and I layers, and photoemission studies reveal an exceptionally large Rashba parameter $\alpha$ \cite{ishizaka2011giant}.  In GeTe, a polar distortion of the rhombohedral unit cell leads to inversion asymmetry and ferroelectricity \cite{liebmann2016giant,krempasky2015surface,kolobov2014ferroelectric}.  Both materials are semiconductors in which the valence and conduction bands are described by an effective Rashba model with symmetry-breaking direction $\hat n$, which is determined by the crystal structure.  There is interest in finding other ferroelectric materials with strong spin-orbit coupling, motivated by the desire to control the direction of $\hat n$ with an applied electric field \cite{di2013electric,narayan2015class,stroppa2014tunable,kepenekian2015rashba,kim2014switchable,kolobov2014ferroelectric}.

In this work, we study the influence of a 3-d (nonmagnetic) Rashba material on the magnetic anisotropy of an adjacent ferromagnetic layer.  The interface between these materials breaks the symmetry along the $\hat z$-direction, and the addition of another symmetry breaking direction $\hat n$ enriches the magnetic anisotropy energy landscape.  For a general direction of $\hat n$, we find a complex dependence of the system energy on magnetic orientation.  In our model system, we find the out-of-plane anisotropy is much smaller than the demagnetization energy.  However an in-plane component of $\hat n$ leads to a uniaxial in-plane magnetic anisotropy which can be on the order of (or larger than) the magnetocrystalline anisotropy of bulk ferromagnetic materials.  Control of $\hat n$ (for example via an electric field in a 3-d Rashba ferroelectric) can therefore modify the magnetization orientation, opening up new routes to magnetic control.

\begin{figure}[h!]
\includegraphics[width=0.46\textwidth]{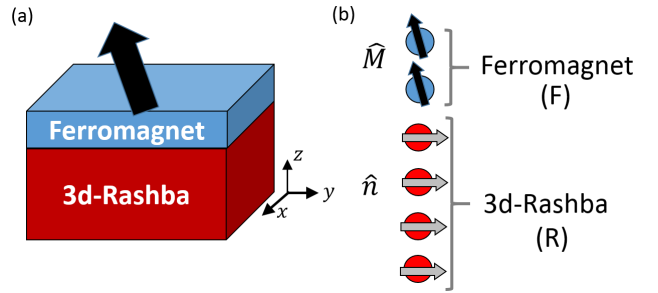}
\caption{(a) shows the system geometry of a thin film ferromagnet adjacent to a nonmagnetic material described by a 3-d Rashba model. (b) shows the unit cell of the model system.}\label{fig:geom}
\end{figure}

\section{Numerical evaluation of 2-d model}

We first consider a bilayer system as shown in Fig. \ref{fig:geom}(a), with unit cell as shown in Fig. \ref{fig:geom}(b).  We take 4 layers of the Rashba material and 2 layers of the ferromagnetic material with stacking along the $z$-direction, and assume a periodic square lattice in the $x-y$ plane with lattice constant $a$.  The Hamiltonian of the system is given by $H=H_{\rm TB}+H_{\rm F}+H_{\rm R}+H_{\rm F-R}$, where:
\begin{eqnarray}
H_{\rm TB} &=& t \sum_{\langle ij\rangle} c_i^\dagger c_j \label{eq:htb} \\
H_{\rm F} &=&\frac{\Delta}{2} \sum_{i\gamma\nu} c_{i\gamma}^\dagger c_{i\nu} \left(\hat M \cdot \vec \sigma_{\gamma\nu}\right) +  H_{\rm TB} ,\label{eq:hf}  \\
H_{\rm R} &=&i \frac{\alpha}{2a}\sum_{\langle ij\rangle\gamma\nu} c_{i\gamma}^\dagger c_{j\nu} \left[\left(\hat{r}_{ij} \times \vec{\sigma}_{\gamma\nu}\right)\cdot \hat n\right] +  H_{\rm TB} , \label{eq:hr}\\
H_{\rm F-R} &=& H_{\rm TB} + \frac{\alpha}{2a}\left( i c_{F\gamma}^\dagger c_{R\nu} \sigma^x_{\gamma\nu} + {\rm h.c.}\right) \label{eq:hcouple} .
\end{eqnarray}
$H_{\rm TB}$ describes nearest-neighbor hopping with amplitude $t$. $H_{\rm F}$ is the on-site spin-dependent exchange interaction in the ferromagnet.  Its magnitude is $\Delta$ and is directed along the magnetization orientation $\hat M$.  $H_{\rm R}$ describes the Rashba layer: spin-orbit coupling and the broken symmetry direction $\hat n$ lead to spin-dependent hopping between sites $i$ and $j$ which is aligned along the $\hat{r}_{ij}\times{\hat n}$ direction, where ${\hat r}_{ij}$ is the direction of the vector connecting sites $i$ and $j$.  $\alpha$ is the Rashba parameter (with units of energy$\times$ length).  $H_{F-R}$ is the coupling between ferromagnetic and Rashba layers - it includes both spin-independent hopping and spin-dependent hopping.  In Eq. \ref{eq:hcouple}, the $F$ ($R$) subscript in the creation and annihilation operators labels the interfacial ferromagnet (Rashba) layer.  We find the model results are similar if $H_{\rm F-R}$ includes only spin-independent hopping.

The Fermi energy $E_F$ is determined by the electron density $\rho$ and temperature $T$ according to:
\begin{eqnarray}
\rho &=& \frac{1}{\left(2\pi\right)^2} \int d{\bf k}~ f\left(\frac{ E_{\bf k}-E_F}{k_{\rm B}T}\right)\label{eq:n}
\end{eqnarray}
where $k_{\rm B}$ is the Boltzmann constant, and $f\left(x\right)$ is the Fermi distribution function: $f\left(x\right) = \left(1+e^{x}\right)^{-1}$.  The integral is taken over the two-dimensional Brillouin zone.  For a given electron density $\rho$, Eq. \ref{eq:n} determines the Fermi energy (which generally depends on $\hat{M}$).  The total electronic energy is then given by:
\begin{eqnarray}
E(\hat M) &=& \frac{1}{\left(2\pi\right)^2} \int d{\bf k}~ E_{\bf k} f\left(\frac{ E_{\bf k}-E_F}{k_{\rm B}T}\right)\label{eq:e}
\end{eqnarray}

The default parameters we use are $\Delta=t,~\alpha=0.4\times ta$.  For $t=1~{\rm eV}$ and $a=0.4~{\rm nm}$, this corresponds to a Rashba parameter of $0.16~{\rm eV\cdot nm}$ (compared to a value $\alpha=0.38~{\rm eV\cdot nm}$ for BiTeI \cite{ishizaka2011giant}).  We let $T=0.1~{\rm K}$ and use a minimum of $1200^2~{\bf k}$-points to evaluate the integrals in Eqs. \ref{eq:n}-\ref{eq:e}.

\begin{figure}[h!]
\includegraphics[width=0.46\textwidth]{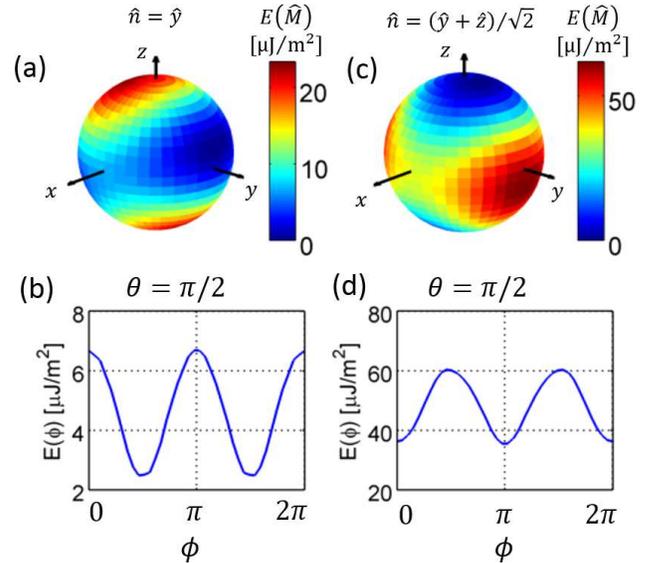}
\caption{(a) The system energy as a function of the magnetic orientation for $\hat n=\hat y$, (c) shows the same for $\hat n=\left(\hat y + \hat z\right)/\sqrt{2}$.  (b) shows the system energy versus magnetic orientation within the $x-y$ plane, parameterized by the azimuthal angle $\phi$ for $\hat n=\hat y$ (d) shows the same for $\hat n=\left(\hat y + \hat z\right)/\sqrt{2}$. }\label{fig:sphere}
\end{figure}

Fig. \ref{fig:sphere}(a) shows the total energy versus magnetic orientation for $\hat{n}=\hat{y}$.  We observe an out-of-plane magnetic anisotropy, however its magnitude is much smaller than the demagnetization energy, which is typically on the order of $1000~{\rm \mu J/m^2}$.  In the rest of the paper, we assume that the demagnetization energy leads to easy-plane anisotropy of the ferromagnet, so that the most relevant features of the Rashba substrate-induced anisotropy energy are confined to the $x-y$ plane.  The energy versus easy-plane magnetic orientation (parameterized by the azimuthal angle $\phi$) is shown in Fig. \ref{fig:sphere}(b).  The anisotropy is uniaxial and favors orientation in the $\pm \hat y$-directions.  This is in contrast to the bulk magnetocrystalline anisotropy of cubic transition metal ferromagnets, which has in-plane biaxial anisotropy.

As a point of reference for the magnitude of the calculated substrate-induced uniaxial anisotropy, we compare it to the magnetocrystalline anisotropy $E_{{\rm MC}}$ for 2-monolayer thick film of Fe, Ni, and Co, for which $E_{\rm MC}$=$\left(34,~3.5,~318\right)~{\rm \mu J/m^2}$, respectively \cite{cullity2011introduction}.  Fig. \ref{fig:sphere}(b) shows that for the material parameters in our model, the Rashba-induced uniaxial anisotropy is larger than the magnetocrystalline anisotropy of Ni.  We also note that permalloy, commonly used as a thin film ferromagnet, has a vanishing magnetocrystalline anisotropy \cite{yin2006magnetocrystalline}.

Next we consider Rashba layer with both in-plane and out-of-plane components of $\hat n$: $\hat n = \left(\hat y + \hat z\right)/\sqrt{2}$.  This is motivated in part by the fact that the symmetry is strongly broken along the $\hat z$-direction by the interface.  Our interest is in the influence of an out-of-plane component $\hat n$  when there is also an in-plane component of $\hat n$.  The resulting $E(\hat M)$ shown in Fig. \ref{fig:sphere}(c).  As before, there is an in-plane uniaxial anisotropy, as shown in Fig. \ref{fig:sphere}(d), with a larger energy barrier as the previous $\hat n=\hat y$ case.  Note that the $\hat y$ direction is now a hard axis.  In general we find that for an in-plane component of $\hat n$, the $\hat n\times \hat z$ direction can be either a hard or an easy axis, depending on details of the electronic structure.

\begin{figure}[h!]
\includegraphics[width=0.48\textwidth]{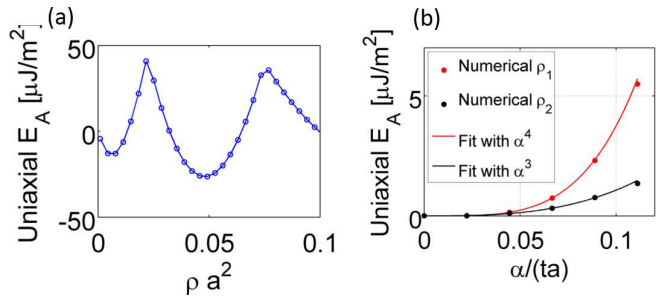}
\caption{(a) $E_A$ versus dimensionless electron density $\rho a^2$.  (b) $E_A$ versus $\alpha$ for two values of $\rho$ ($\rho_1 a^2=0.02,\rho_2 a^2=2.1$), along with fitting to powers of $\alpha$.  For small $\rho$, the $\alpha^4$ dependence can be understood in terms of perturbation theory.}\label{fig:params}
\end{figure}

We define the uniaxial in-plane anisotropy energy $E_A$ as the difference in energy for $\hat M=\hat y$ and $\hat M=\hat x$.  Fig. \ref{fig:params}(a) shows $E_A$ as the electron density $\rho$ is varied, and indicates that sign of the anisotropy can change depending on the value of $\rho$.  Fig. \ref{fig:params}(b) shows the dependence of $E_A$ on $\alpha$ for two values of the electron density $\rho$.   We find that the uniaxial anisotropy energy varies as a power of $\alpha$ which depends on $\rho$ (or equivalently on the band filling).  The origin of this dependence is discussed in more detail in the analytic model we develop next.

\section{Analytical treatment of 1-d model}
\begin{figure}[h!]
\includegraphics[width=0.46\textwidth]{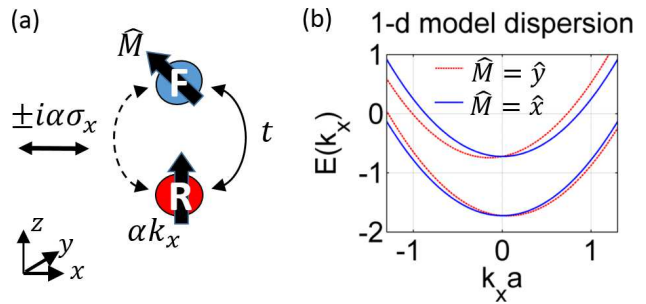}
\caption{(a) cartoon of the unit cell of a 1-d model which extends along the $x$-direction.  The top site is ferromagnetic and the bottom site includes Rashba spin-orbit coupling.  The arrow along the $+\hat{z}$-direction on the R site indicates the direction of the Rashba effective magnetic field (we've assumed $k_x>0$ in the figure).  The double arrow along the $\pm x$-direction depicts the spin direction of the spin-dependent hopping between F and R sites.  (b)  The energy dispersion for the lower two bands for two magnetic orientations ($\Delta=1,~\alpha=0.5$ for this plot).}\label{fig:1d}
\end{figure}

To gain some insight into the physical origin of the uniaxial in-plane anisotropy, we consider a simplified system of a 1-d chain of atoms extending in the $x$-direction with 2 sites in the unit cell, as shown in Fig. \ref{fig:1d}(a), and take $\hat n=\hat y$.   The Hamiltonian is the same as Eqs. \ref{eq:htb}-\ref{eq:hcouple}.  We compute $E(k_x)$ in a perturbation expansion of the spin-orbit parameter $\alpha$, then determine the total energy as a function of $\alpha$ and $\hat M$.

The lowest order term in $\alpha$ is of the form $E(k_x)\propto \alpha k_x M_z$.  This can be understood as a simple magnetic exchange interaction between the F and R sites: spin-orbit coupling leads to effective magnetic field on R along the $z$-direction.  This effective magnetic field is proportional to $\alpha k_x$ and is exchange coupled to the effective magnetic field on the F site (see Fig. \ref{fig:1d}(a)).  The linear-in-$k_x$ term in $E(k_x)$ shifts the energy bands downward by an amount proportional to the square of the coefficient multiplying $k_x$.  The total energy decreases by the same factor.  This finally results in a magnetic anisotropy energy which is proportional to $\alpha^2 M_z^2$, describing an out-of-plane anisotropy.



As discussed earlier, we assume the ferromagnet is easy-plane and are therefore interested in the magnetic anisotropy within the plane.  In-plane anisotropy in $E(k)$ appears only at 2nd order in $\alpha$.  Taking $M_z=0$, we find $E(k)$ takes a simple form in the limits that $\alpha \ll \Delta \ll 2t$ and $ka\ll 1$.  We present the result for the lowest energy band:
\begin{eqnarray}
E\left(k_x\right) &=& A k_x^2 + B\alpha^2 k_x M_y  + C \label{eq:ek}
\end{eqnarray}
Where $A,~B,~C$ depend on $\Delta$ and $t$, whose precise form is not essential for this discussion \cite{footnote}.  Fig. \ref{fig:1d}(b) shows the numerically computed dispersion for two orientations of the ferromagnet.  For $\hat M=\hat x$, the energy bands are purely quadratic in $k_x$, while for $\hat M=\hat y$, the energy bands acquire a linear-in-$k_x$ component, which is of opposite sign for the two lowest energy bands.  In the case where only the lowest band is occupied, the total energy again depends on the square of the linear-in-$k_x$ coefficient, resulting in an in-plane anisotropy energy which is proportional to $\alpha^4 M_y^2$.

We can understand the physical origin of the dependence of the energy on $M_y$:  the spin-dependent hopping between F and R sites induces a twisted exchange interaction between the spins on these sites, which favors a noncollinear configuration in which both spins lie in the $y-z$ plane \cite{imamura2004twisted}.  The effective magnetic field on the R-site is always along the $z$-direction, so the exchange energy therefore differs when the ferromagnet is aligned in the $x$ versus $y$ direction.  The twisted exchange interaction energy contains a factor proportional to the effective magnetic field on R (which varies as $\alpha k_x$), and a factor proportional to the spin-dependent hopping (which is linear in $\alpha$) - so the energy varies as finally as $\alpha^2 k_x M_y$.  We find that the lowest energy band of the 2-d system of the previous section also contains a linear-in-$k_x$ term which varies as $\alpha^2 k_x M_y$, indicating that the physical picture developed for the 1-d system also applies for the 2-d system.  In the case where multiple bands are occupied, we're unable to find a closed form solution for the anisotropy energy, and find that it can vary with a power of $\alpha$ that depends on the electron density (and corresponding Fermi level).  This is shown in Fig. \ref{fig:params}(b) where the uniaxial anisotropy varies as $\alpha^3$ for multiply filled bands (we've observed several different power-law scalings with $\alpha$ for different system parameters).  Nevertheless the case of a singly occupied band is sufficient to illustrate the physical mechanism underlying the in-plane magnetic anisotropy.


\section{Conclusion}

We've examined the influence of a 3-d Rashba material on the magnetic properties of an adjacent ferromagnetic layer.  A uniaxial magnetic anisotropy is developed within the plane of the ferromagnetic layer with easy-axis direction determined by $\hat n$.  Depending on material parameters, the easy-axis can be parallel or perpendicular to $\hat n$. For large but realistic values of the bulk Rashba parameter of the substrate, the magnitude of the anisotropy indicates that the effect should be observable.  For materials in which the direction of the bulk symmetry breaking is tunable - for example in a ferroelectric 3-d Rashba material - the interfacial magnetic anisotropy offers a novel route to controlling the magnetic orientation.  The magnitude of the Rashba-induced anisotropy is much less than the demagnetization energy, so its influence is confined to fixing the in-plane component of $\hat M$.  This control would nevertheless be useful in a bilayer geometry in which electrical current flows in plane.  In this case the anisotropic magnetoresistance effect yields a resistance which varies as $(\hat J\cdot \hat M)^2$ \cite{mcguire1975anisotropic}, which can be utilized to read out the orientation of $\hat M$.  We also note recent works have utilized the reduced crystal symmetry of the substrate to achieve novel directions of current-induced spin-orbit torques \cite{ralph}.  This indicates that the symmetry of the substrate can influence the nonequilibrium properties of the magnetization dynamics, in addition to modifying the equilibrium magnetic properties, as studied in this work.

\section*{Acknowledgment}
J. Li acknowledges support under the Cooperative Research Agreement between the University of Maryland and the National Institute of Standards and Technology Center for Nanoscale Science and Technology, Award 70NANB10H193, through the University of Maryland.

\bibliographystyle{apsrev}
\bibliography{ref}

\begin{thebibliography}{28}
\bibitem[{foo()}]{footnote}
\bibinfo{note}{For completeness we give the forms for the lowest two energy
  bands here, $A_{1,2}=1 - \alpha^2/64\left(\pm 16 + 8\Delta \mp 16 \Delta^2 +
  7\Delta^3\right)$, $B_{1,2}=\left(\mp1 + \Delta/2\right)$, and
  $C_{1,2}=1/16\left(-8\alpha^2+\Delta^2\alpha^2-16\mp8\Delta-\Delta^2\right)$.
  The $\alpha$-dependent terms in these factors lead to higher order
  corrections to the total energy.}
\end{thebibliography}

\end{document}